# Spin-Induced Linear Polarization of Excitonic Emission in Antiferromagnetic van der Waals Crystals

Xingzhi Wang,[1][*] Jun Cao,[1] Zhengguang Lu,[2,3] Arielle Cohen,[4] Hikari Kitadai,[1] Tianshu Li,[4] Matthew Wilson,[5] Chun Hung Lui,[5] Dmitry Smirnov,[2] Sahar Sharifzadeh,[1,4,6,7] Xi Ling[1,4,8][*]

[1] Department of Chemistry, Boston University, Boston, Massachusetts 02215, USA.

[2] National High Magnetic Field Laboratory, Tallahassee, Florida 32310, USA.

[3] Department of Physics, Florida State University, Tallahassee, Florida 32306, USA.

[4] Division of Materials Science and Engineering, Boston University, Boston, Massachusetts 02215, USA.

[5] Department of Physics and Astronomy, University of California, Riverside, California 92521, USA

[6] Department of Electrical and Computer Engineering, Boston University, Boston, Massachusetts 02215, USA.

[7] Department of Physics, Boston University, Boston, Massachusetts 02215, USA.

[8] The Photonics Center, Boston University, Boston, Massachusetts 02215, USA.

[*]To whom the correspondence should be addressed. Email address: xzwang@bu.edu; xiling@bu.edu

**Antiferromagnets display enormous potential in spintronics owing to its intrinsic nature, including terahertz resonance[1,2], multilevel states[3,4], and absence of stray fields[5,6]. Combining with the layered nature, van der Waals (vdW) antiferromagnets hold the promise in providing new insights and new designs in two-dimensional (2D) spintronics. The zero net magnetic moments of vdW antiferromagnets strengthens the spin stability, however, impedes the correlation between spin and other excitation elements, like excitons[7,8]. Such coupling is urgently anticipated for fundamental magneto-optical studies and potential opto-spintronic devices. Here, we report an ultra-sharp excitonic emission with excellent monochromaticity in antiferromagnetic nickel phosphorus trisulfides ($NiPS_3$) from bulk to atomically thin flakes. We prove that the linear polarization of the excitonic luminescence is perpendicular to the ordered spin orientation in $NiPS_3$. By applying an in-plane magnetic field to alter the spin orientation, we further manipulate the excitonic emission polarization. Such strong correlation between exciton and spins provides new insights for the study of magneto-optics in 2D materials, and hence opens a path for developing opto-spintronic devices and antiferromagnet-based quantum information technologies.**

Since the discovery of monolayers ferromagnets in 2017, magneto-optics plays a compelling role in revealing new physics of magnetism in the extreme nanoscale limit[9-11]. For example, the spin orientation in 2D ferromagnets can be detected by using the magneto-optical Kerr effect (MOKE)[9,10] or spontaneous circularly polarized photoluminescence (PL)[12]. Compared to ferromagnets, however, the studies of magneto-optics in 2D antiferromagnets are far more difficult due to the lack of net magnetic moments. Prior researchers have applied Raman spectroscopy[13-15] and second harmonic generation (SHG)[16,17] to detect the symmetry of spin order in 2D antiferromagnets, based on the expansion of unit cell and centrosymmetry breaking. Nevertheless, these methods are not sufficient to provide direct information of spin properties in 2D magnets, such as the local spin orientation.

Owing to the reduced dielectric screening and enhanced Coulomb interactions, 2D semiconductors exhibit significant excitonic effects, which dominate the optical and optoelectronic properties of materials[18,19]. Particularly, excitonic states in 2D transition metal dichalcogenides (TMDs) exhibit distinctive spin-valley configurations for novel applications of spintronics and valleytronics[20-22]. Such phenomenon holds promise for investigating the spin properties through excitons and controlling the optoelectronic properties by spins, which provides new concepts in the design of novel opto-spintronics. However, the fact that TMDs are not intrinsically magnetic restricts the realization of spin-exciton coupling in specific conditions, which diminishes new insights in the fundamental physics and limits the response wavelength range for opto-spintronic device. Although theoretical calculations have predicted excitonic transitions below the band gap of vdW magnets[23], direct experimental evidence of excitonic emission in ultrathin magnetic materials is still lacking.

Here, we report the direct observation of excitonic photoluminescence in the vdW antiferromagnetic $NiPS_3$ from bulk to trilayer flakes. The excitonic emission exhibits excellent monochromaticity with a near-intrinsic linewidth (~330 μeV at temperature T ~ 5 K) in the near-infrared range (with photon energy ~1.476 eV). Notably, the excitonic emission exhibits a linear polarization induced by the anisotropic spin orientations. Such spin-exciton coupling allows us to easily read the spin orientation in bulk and atomically thin $NiPS_3$ flakes by steady-state PL spectroscopy. We further manipulate the spin orientation and the resultant change of exciton polarization by applying an in-plane magnetic field. Our studies demonstrate the strong correlation between exciton and ordered spins in atomically thin antiferromagnets, and paves the way for developing new opto-spintronic devices in 2D systems.

Monolayer $NiPS_3$ crystal has hexagonal lattice structure with three-fold rotation symmetry. The material properties hence exhibit distinctive in-plane isotropy (Fig. 1a)[24]. Such in-plane isotropy is largely retained in the monoclinic bulk crystal due to the weak interlayer coupling, though the interlayer stacking order in principle breaks the three-fold rotation symmetry[15,24]. For example, bulk $NiPS_3$ exhibits similar isotropic Raman response as the monolayers[15]. However, when $NiPS_3$ transits from the paramagnetic phase to the antiferromagnetic phase below its Néel temperature ($T_N$ ~152 K), the spins are aligned either in parallel or antiparallel along the in-plane $a$-axis with a small out-of-plane component as shown in Fig. 1b[25]. In each single layer, both the spin orientation and ferromagnetic chains distinguish the *a*-axis from the other two directions rotated by ±120° and hence breaks the three-fold rotation symmetry.

To explore these antiferromagnetic spins, we have measured the PL of high-quality NiPS$_3$ single crystals and exfoliated flakes (Fig. 1c and Supplementary Fig. 1 and 2). Fig. 1d shows the typical PL spectra of bulk and trilayer NiPS$_3$ under 568-nm continuous-wave laser excitation at T ~ 5 K. A distinctly sharp peak (labeled as X) is resolved at ~1.476 eV in the spectra of both the bulk and trilayer samples. The trilayer PL peak is only slightly (~0.8 meV) higher than the bulk PL peak, and both are close to the band gap energy (~1.5 eV) of NiPS$_3$[24,26]. Here we will focus on peak X, while other nearby emission peaks are discussed in the Supplementary Information. We will first show that peak X originates from the excitons and then demonstrate its correlation with the spin orientation.

We attribute peak X to the excitons in NiPS$_3$ for a few reasons. First, we observe an absorption peak at ~ 1.480 eV, only ~ 4 meV above the PL peak X at ~ 1.476 eV, strongly suggesting the excitonic origin of peak X. Also, the two peaks exhibit similar temperature dependence (Fig. 2a), suggesting that they share the same origin. The noticeable absorption also excludes defect states as the origin of the emission, because defect absorption is generally invisible in crystals due to the limited density of states. Second, the PL intensity of peak X increases almost linearly with the excitation laser power (Supplementary Fig. 4b), in contrast to the saturation behavior expected in defect emission[27]. Third, peak X is not a Raman mode because its photon energy remains unchanged under laser excitation at 458 and 568 nm (Supplementary Fig. 4a), whereas Raman modes should shift with the excitation energy. Finally, peak X disappears at T > 120 K, consistent with exciton dissociation at high temperature (Fig. 2b).

Peak X exhibits an exceptionally sharp PL line width of only ~330 μeV (the inset in Fig 1d). Such a narrow line width strongly suggests that the emission comes from bound excitons, rather than free-particle interband transitions which typically exhibit a broad emission band. As the temperature increases, both the absorption and emission peaks redshift and broaden (Fig. 2a and 2b) and disappear at T ~ 120 K, presumably due to exciton dissociation (Supplementary Fig. 5). The temperature-dependent integrated PL intensity, full width at half maximum (FWHM), and peak position of peak X exhibit gradual variations as the temperature decreases from 120 to 20 K (Fig. 2c and Supplementary Fig. 4c). These behaviors are distinct from most phenomena induced by magnetic phase transition, which usually exhibit a non-differentiable inflection at $T_N$[13-17]. In addition, the temperature at which peak X appears (~120 K) is lower than the $T_N$ of NiPS$_3$ (~152 K). These results suggest that the appearance of peak X does not originate from the magnetic transition.

We have calculated the electronic band structure of bulk NiPS$_3$ using the density functional theory (DFT), including a Hubbard U parameter on the Ni atoms (Supplementary Fig. 7). The calculated band structure suggests an indirect band gap of 1.55 eV in bulk NiPS$_3$. However, the observation of peak X in the absorption spectrum and the small Stokes shift (~ 4 meV) suggest that this exciton is related to the direct transition rather than the indirect band gap transition. This contradiction can be explained by the predicted unique near-flat electronic bands near the conduction band minimum (CBM) of NiPS$_3$, due to the localized Ni $3d$ orbitals (Supplementary Fig. 7). These localized states in the conduction bands result in a more likely direct transition at the valence band maximum (VBM) compared to the indirect transition from VBM to CBM (see Supplementary Information). Therefore, we assign peak X to the direct transition from VBM to the lowest near-flat conduction band and the broad PL peak at the lower energy side of the X peak to the emission caused by the indirect transition (Fig. 2d). The corresponding predicted direct transition energy ~ 1.7 eV is larger than the experimental energy of peak X (1.476 eV). This discrepancy may be due to the neglect of excitonic effects and the error of band gap calculation due to the use of an approximate U value in the DFT+U method (see Supplementary Information). Due to the large electron effective mass in the conduction band, the exciton effective mass approximately equals to the hole effective mass in the dispersive VBM that consists mainly of S $p$-orbitals (Supplementary Fig. 7c). The exciton binding energy is estimated to be tens of meV according to the thermal energy of the temperature (~120 K) at which the excitons dissociate. The small binding energy indicates the Wannier-like nature of the excitons. We have further measured the exciton lifetime by using time-resolved PL spectroscopy (Fig. 2e). The exciton lifetime, after deconvolution with the instrument response function, is estimated to have an upper bound of ~10 ps. Such a short lifetime is normally considered non-radiative due to the existence of defects in the crystals[28].

After elucidating the excitonic origin of peak X, we turn to study its correlation with the spin order. According to our DFT calculations, the low-lying near-flat conduction bands are mainly composed of Ni $d$-orbital electrons (Supplementary Fig. 7c). Since the $d$-orbit electrons contribute predominantly to the magnetic moment of $NiPS_3$, we hypothesize that the associated excitons could couple to the spins in the material. To this end, we have measured the polarization-dependent PL in antiferromagnetic $NiPS_3$ in two configurations (Supplementary Fig. 10). In the first configuration, we rotate the incident laser polarization ($P_{in}$) and measure the total PL intensity of peak X with a fixed collection polarization ($P_{col}$). We do not observe any dependence of the PL on $P_{in}$ (Fig. 3a). As the incident photon energy (~2.54 eV) far exceeds the $NiPS_3$ band gap (~1.5 eV), the carriers are insensitive to the excitation photon polarization after relaxing from the high-energy initial states. This accounts for the isotropy of the optical absorption and resultant PL.

In the second configuration, we maintain the same excitation laser polarization, but collect PL of peak X at different polarization ($P_{col}$). Remarkably, we find that the collected PL varies significantly with $P_{col}$ (Fig. 3a). The PL intensity reaches the maximum (minimum) when $P_{col}$ is perpendicular (parallel) to the $a$-axis. From bulk down to few-layer samples, a similar phenomenon is observed and the PL intensity can be fit by a sinusoidal function: $I(\theta) = I_0 + I_1 \sin^2 \theta$. Here $\theta$ is the angle between the PL polarization and the $NiPS_3$ $a$-axis; $I_0$ and $I_1$ are fitting constants (dashed lines in Fig 3a and Supplementary Fig. 11). We determine the $a$- and $b$-axes of our samples by selected-area electron diffraction (SAED) (Supplementary Fig. 12 and Supplementary Information). As the antiferromagnetic spins are parallel to the $a$-axis, our experimental results suggest a correlation between the excitonic emission and the spins (Fig. 3b), with the electric dipole of excitons perpendicular to the spin direction in $NiPS_3$. We define the PL linear polarization degree, $\rho = (I_b - I_a)/(I_b + I_a)$, where $I_a$ ($I_b$) is the PL peak intensity when $P_{col}$ is parallel to $a$-axis ($b$-axis). We find that $\rho$ is ~ 68 % at T < 70 K, and drops significantly with increasing temperature (Fig. 3c). The decreasing $\rho$ at high temperature can be attributed to phonon scattering, which suppresses the coupling between excitons and ordered spins in the materials.

One might argue that the exciton PL polarization is correlated to the interlayer stacking order, which can induce lattice anisotropy between *a*- and *b*-axis. To confirm the spin induced origin of the exciton polarization, we apply an in-plane magnetic field (B) to tune the spin orientation in NiPS$_3$ (Fig. 4a). We mount two bulk NiPS$_3$ samples (labeled as S$_1$ and S$_2$) inside the magnetic cell. Their crystal orientations (i.e. *a*-axis) are rotated by 120º from each other. As a result, the PL polarizations of peak X from sample S$_1$ and S$_2$ are off by 120º at zero magnetic field (Fig. 4b), consistent with our results in Fig. 3a. Then we apply an in-plane magnetic field B up to 25 T along the *a*-axis direction of S$_1$ and conduct the polarized PL measurement (Fig. 4c-4e). The strong magnetic field will cause all the spins in S$_1$ and S$_2$ to point along the same direction as the magnetic field, regardless of the crystal orientation, thus changing the antiferromagnetic order into a ferromagnetic order (Fig. 4f). Remarkably, the S$_1$ and S$_2$ PL show the same polarization at B = 25 T (Fig. 4e). The result unambiguously verifies that the exciton polarization is determined by the spin direction, rather than the weak crystalline anisotropy from the layer stacking. It also shows that the antiferromagnetic order is not directly responsible for the exciton emission.

To further probe the transition from the intrinsic antiferromagnetic order to the field-induced ferromagnetic order, we compare the PL polarization angle $\varphi_1$ ($\varphi_2$) for sample S$_1$ (S$_2$) at different B-field (Fig. 4g). After excluding the effect from Faraday rotation, $\varphi_1$ remains unchanged at ~ 0 degree. In contrast, $\varphi_2$ shifts nonlinearly with the field and abruptly approaches $\varphi_1$ when the B-field is between 10 and 15 T. The spin orientation can be considered as perpendicular to the PL polarization. The material is in the antiferromagnetic phase when $\varphi_2 - \varphi_1 \sim 120º$ at B < 10 T and it transits to the spin-flop phase with ferromagnetic order when $\varphi_2$ is approximately equal to $\varphi_1$ at B > 15 T. Even though the manipulation of exciton was only demonstrated in bulk NiPS$_3$, we anticipate similar results on few-layer flakes, in which the linearly polarized excitonic emission has been demonstrated. Also, the phenomena should persist at least down to the bilayer, due to the unchanged crystal symmetry and magnetic structure as the bulk, but it remains a question whether or not the same phenomenon exists in the monolayer due to the unstable magnetic structure and different crystal symmetry[15].

The strong correlation between excitons and spins in vdW antiferromagnet will greatly benefit future fundamental researches in magnetism and magneto-optics. Most of the prior optical spin-probe techniques in antiferromagnets, which rely on magneto-optical effect (e.g. Faraday effect and Voigt effect), require an ultrafast laser to induce the thermal or magnetic perturbation[29,30], with the amplitude of signals proportion to the thickness of samples[7,30]. In contrast, the spin-induced polarization of excitonic emission demonstrated here offers a steady, non-destructive and sensitive optical method to investigate the spin properties of atomically thin layers. More generally, our experiment also suggests strong spin-exciton correlation in the materials with other magnetic phases, including ferromagnetic orders. The conceptually new mechanism of spin-exciton coupling will stimulate the future theoretical and experimental studies in the field, promoting the development of opto-spintronic device and magnetic quantum information technology.

**Acknowledgements**

X.W. and X.L. acknowledge the financial support from Boston University and the Boston University Photonics Center. The TEM imaging was performed at the Center for Nanoscale Systems (CNS), a member of the National Nanotechnology Coordinated Infrastructure Network (NNCI), which is supported by the National Science Foundation under NSF award no. 1541959. CNS is part of Harvard University. A.C. and S.S. acknowledge financial support from the U.S. Department of Energy (DOE), Office of Science, Basic Energy Sciences (BES) Early Career Program under Award No. DESC0018080. The authors acknowledge the computational resources through the Extreme Science and Engineering Discovery Environment (XSEDE), which is supported by National Science Foundation grant number ACI-1548562; and the National Energy Research Scientific Computing Center, a DOE Office of Science User Facility supported by the Office of Science of the U.S. Department of Energy under Contract No. DE-AC02-05CH11231. Z.L. and D.S. acknowledge support from the U.S. Department of Energy (No. DE-FG02-07ER46451) for National High Magnetic Field Laboratory, which is supported by National Science Foundation through NSF/DMR-1644779 and the State of Florida.

## Methods

**Sample preparation.** NiPS$_3$ single crystals were grown using a chemical vapor transport (CVT) method[24]. A stoichiometric amount of high-purity elements (mole ratio Ni:P:S = 1:1:3, around 1 g in total) and iodine (about 10−20 mg) as a transport agent were sealed into a quartz ampule and kept in a two-zone furnace (650−600 °C). The length of the quartz ampule is about 16 cm with a 13 mm external diameter. The pressure inside the ampule was pumped down to $1\times10^{-4}$ Torr. After 1 week of heating, the ampule was cooled down to room temperature with bulk crystals in the lower temperature end. The purity of the synthesized sample has been confirmed by series of characterizations. The powder X-Ray diffraction (PXRD) data was collected on a Bruker D8 Discover diffractometer with Cu K$_\alpha$ radiation of λ = 1.5418 Å at 40 kV and 40 mA. Elemental analysis experiments were conducted using energy dispersive spectroscopy (EDS) attached to a field emission scanning electron microscope (FESEM, Zeiss Supra 55). TEM measurements were performed on a FEI Tecnai Osiris transmission electron microscope, operating at a 200 keV accelerating voltage. SAED were measured on a JEOL 2100 transmission electron microscope. The SAED simulation was perform through STEM_CELL software (49).

Few-layer NiPS$_3$ flakes were prepared on Si substrates with 285-nm SiO$_2$ layer by mechanical exfoliation from a bulk single-crystal. The morphology and thickness of exfoliated flakes were characterized using an optical microscope (Nikon DS-Ri2) and AFM (Bruker Dimension 3000) in a tapping mode.

**Raman scattering, PL and absorptance spectroscopy.** Optical measurements were carried out on a micro-Raman spectrometer (Horiba-JY T64000) and the signal was collected through a 50× long-working-distance objective. A cryostat (Cryo Industry of America, USA) was used to provide a vacuum environment and a continuous temperature from 5 to 300 K by liquid helium flow. Raman scattering measurement was perform using a triple-grating mode with 1800 g/mm gratings, and a 568-nm laser line from a Kr$^+$/Ar$^+$ ion laser (Coherent Innova 70C Spectrum) was used to excite the sample. Micro-PL and micro-absorption measurements were carried out using a single-grating mode. For PL, 458- and 568-nm laser lines from the Kr$^+$/Ar$^+$ ion laser (Coherent Innova 70C Spectrum) were used to excite the sample. The backscattered signal was dispersed with a 150 g/mm or 1800 g/mm grating. A stabilized tungsten-halogen white-light source (SLS201L, Thorlabs) was used to irradiate the sample from the bottom in micro-absorption measurements. The absorptance spectra were determined by $A(\lambda) = 1 - T(\lambda)/T_0(\lambda)$, where $T(T_0)$ is the intensity of light transmit through the sample (substrate).

**Polarization-dependent PL and magneto-PL measurement.** The polarization-dependent PL and magneto-PL experiment were performed in the National High Magnetic Field Lab (NHMFL). In the polarization-dependent PL measurement, a 488-nm laser was used to excite the bulk crystal and exfoliated flakes. A 50× long-working-distance objective was used to focus the excitation onto the sample and collect the PL signal. The magneto-optical experiments were performed with 31 T DC magnet on bulk crystal. A 532-nm continuous-wave laser was used to excite the sample. The sample was vertically put inside the magnetic cell with the surface parallel to the applied magnetic field. A mirror was set between objective and sample with 45º to change the optical path by 90º. A 10× objective was used to focus excitation onto the sample and collect the PL signal. For both the polarization-dependent and magnetic-field-dependent experiment, the PL signal was filtered by an analyzer, subsequently collected by a multi-mode optical fiber and measured by a spectrometer with a CCD camera (Princeton Instruments, IsoPlane 320).

**Time-resolved PL measurement.** Time-resolved PL experiment was conducted using a time-correlated single photon counting technique. The excitation laser pulses come from an oscillator (Light Conversion Inc., Pharos) with 1030 nm output wavelength, 80 MHz repetition rate, and 90 fs pulse duration. The second harmonic (515 nm wavelength) of the laser was used to excite the samples. The PL signal was spectrally dispersed and filtered by a grating monochromator before detection by a photodiode.

**DFT calculation.** DFT calculations were performed using the Vienna Ab Initio Simulation Package (VASP)[31-34]. We utilized the local density approximation (LDA) with a Hubbard U parameter of 4.0 eV on each Nickel atom[24, 26, 35] to describe the correlated behavior of the d electrons, using the simplified, rotationally invariant Dudarev approach implemented in VASP[36]. Projector augmented wave (PAW) potentials describe the core and nuclei of atoms, with 10, 5 and 6 electrons treated explicitly for Ni, P and S respectively[37, 38]. The k-point mesh for self-consistent field and structural optimization was 4×7×5, sufficient to converge the total energy to 1 meV/atom. Additionally, for structural optimization we used a cutoff energy for the planewave basis set of 470 eV and a threshold of 1 meV/Angstrom for all forces. To determine the magnetic state of the NiPS$_3$, we initialized the system with two types of antiferromagnetic and one ferromagnetic state, with an initial magnetic moment of 3.0 or -3.0 assigned to each nickel atom and determined the lowest energy structure. In agreement with previous studies, we found that antiferromagnetically coupled "zigzag" stripes of aligned spins described in Fig. 1b were the lowest energy structure[39]. The other structures considered were a ferromagnetic configuration and an antiferromagnetic configuration in which spins alternated between nearest neighbors.

To calculate the band structure and density of states, we used a k-path between 12 high-symmetry points in the Brillouin zone[40] with 10 k-points along each segment (or 20, when considering spin-orbit coupling). We checked the effect of spin-orbit coupling by performing non-collinear magnetic structure calculations in VASP[41] starting from both the DFT relaxed geometry and the lattice parameters from experiment. In both cases, spin-orbit coupling resulted in narrow band splitting (< 0.01 eV) in some segments of the valance and conduction band.

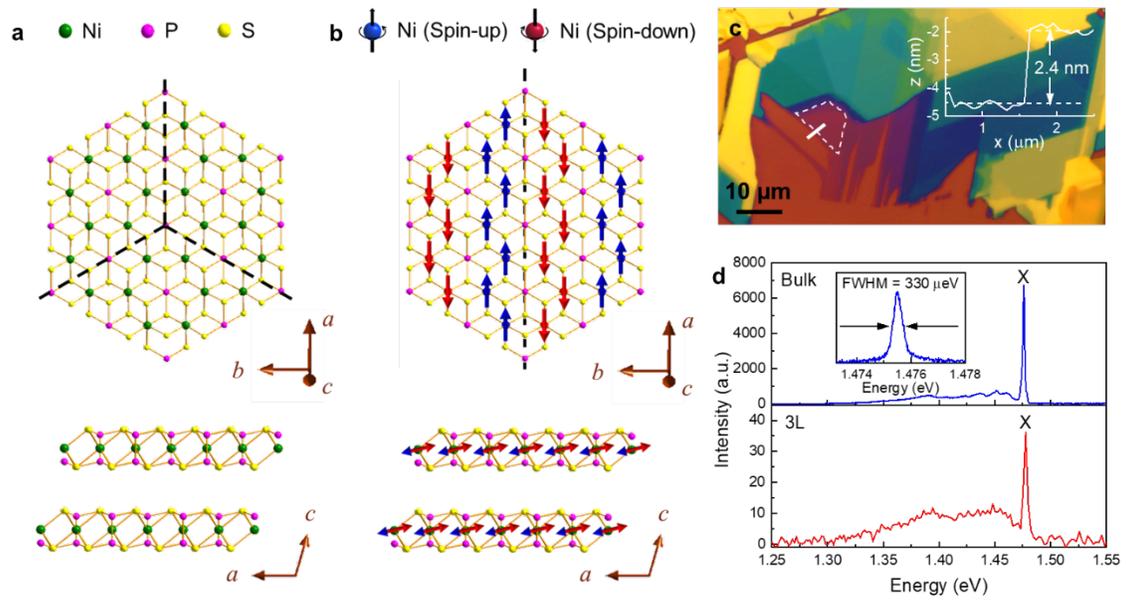

**Fig. 1 | Crystal structure, spin structure, and PL of NiPS$_3$. a**, Top view of the crystal structure of monolayer (up) and side view of the crystal structure of bulk NiPS$_3$ (down). **b**, Top view of the spin structure of monolayer (up) and side view of the spin structure of bulk NiPS$_3$ (down). **c**, Optical microscopy image of exfoliated NiPS$_3$ flakes on a silicon substrate with 285-nm SiO$_2$ thin film. The trilayer flake is marked by the white dashed line and the inset is its height profile along the white solid line in the figure. **d**, PL spectra from bulk and the trilayer NiPS$_3$ flake under 568-nm continuous-wave laser excitation at 5 K. The inset shows a high-resolution spectrum of peak X in bulk NiPS$_3$.

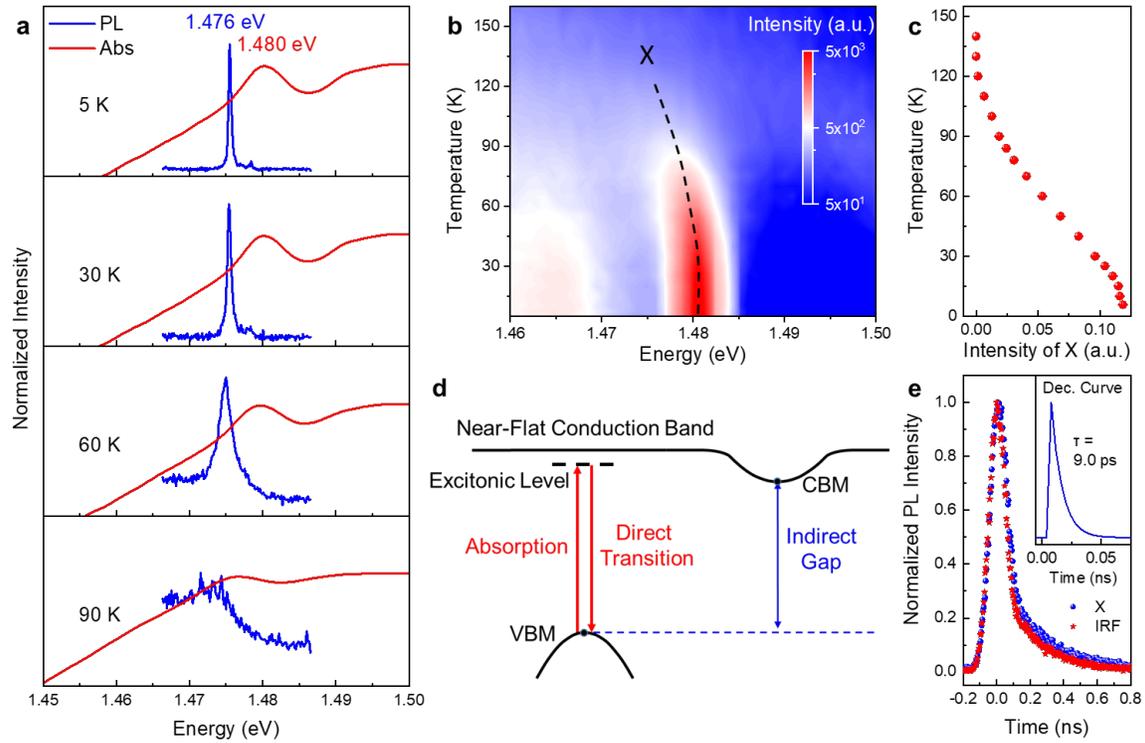

**Fig. 2 | Characterization of excitonic emission in NiPS$_3$. a**, Temperature-dependent PL and absorptance spectra near band edge from 5 to 90 K. **b**, Log-scale temperature-dependent PL color map of peak X under 568-nm laser excitation. **c**, Integrated intensity of peak X as a function of temperature. **d**, Schematic illustration of electronic transition for the origin of peak X. **e**, Time-resolved PL measurement excited by 515-nm pulse laser at 10 K compared with the instrument response function (IRF). The inset is the decay curve of exciton peak after deconvolution with the IRF.

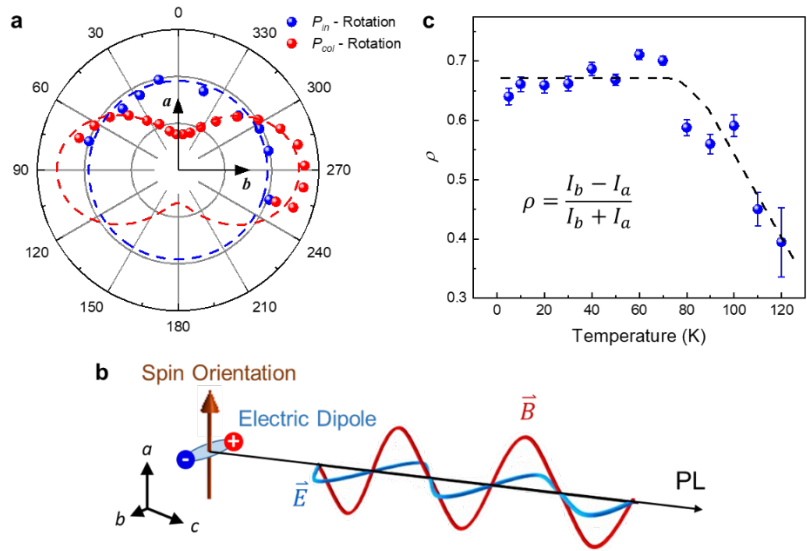

**Fig. 3 | Anisotropic excitonic emission in NiPS$_3$. a**, PL intensity as a function of the incident laser polarization and collection polarization under 488-nm continuous laser excitation at 5 K. **b**, Schematic illustration of polarized PL from antiferromagnetic NiPS$_3$. The electric dipole is perpendicular to the spin orientation. **c**, The PL polarization degree ($\rho = (I_b - I_a)/(I_b + I_a)$) as a function of temperature.

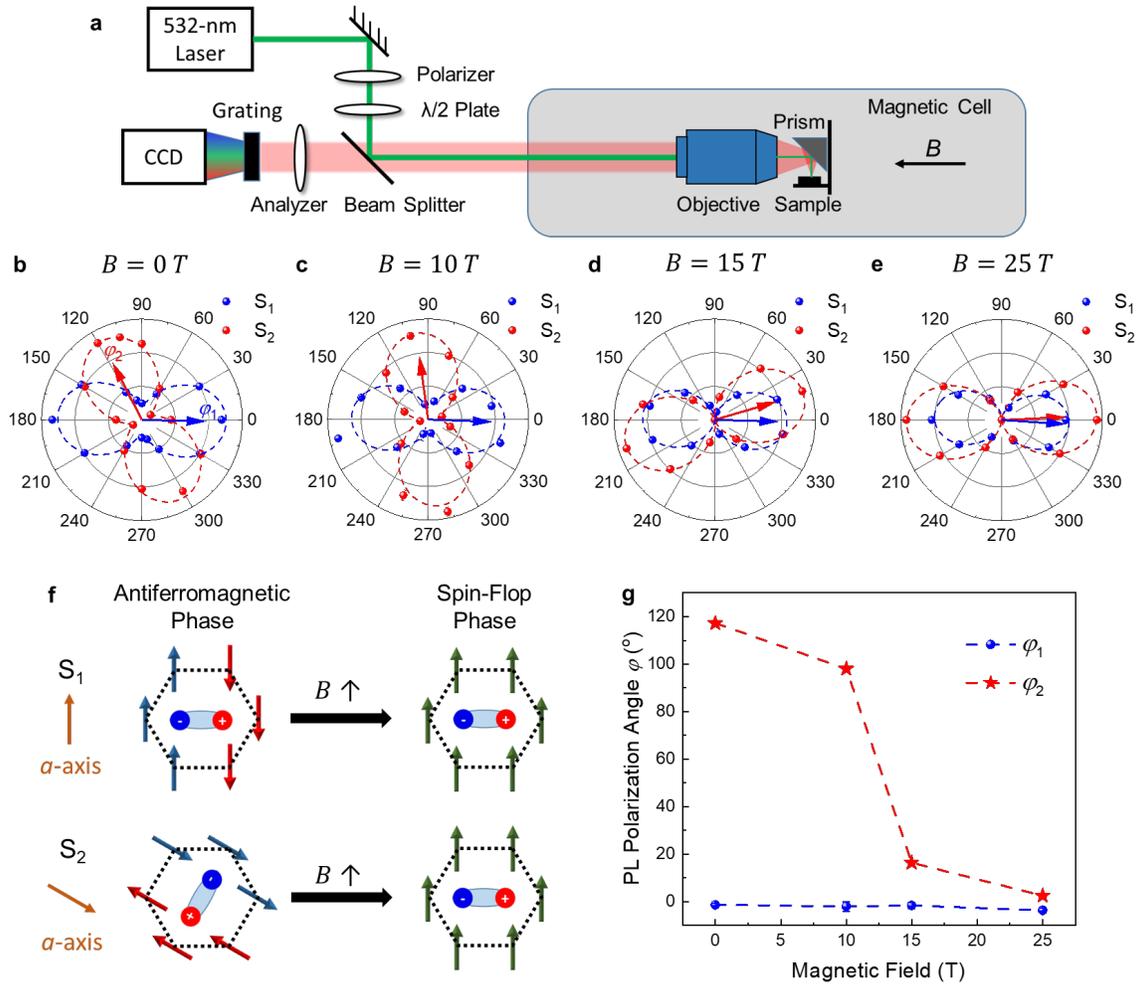

**Fig. 4 | Excitonic detection of spin orientation under in-plane magnetic field. a,** Schematic setup for the magneto-PL experiment in a Voigt geometry. **b-e,** Integrated PL intensity of peak X at (**b**) B = 0, (**c**) B = 10 T, (**d**) B = 15 T, and (**e**) B = 25 T as a function of collection polarization angle on two bulk $NiPS_3$ samples ($S_1$ and $S_2$), whose crystal orientations (i.e. a-axis) are rotated by 120º from each other. The effect from Faraday rotation has been subtracted. **f,** Illustrated diagram of the spin and excitonic dipole orientation at zero and strong in-plane magnetic field. **g,** The PL polarization angle $\varphi_1$ ($\varphi_2$) of sample $S_1$ ($S_2$) as a function of in-plane magnetic field. The results show a transition from the antiferromagnetic phase to spin-flop phase (with ferromagnetic order) from 0 to 25 T.